# Unveiling Vulnerability and Inequality in Disrupted Access to Dialysis Centers During Urban Flooding

*Faxi Yuan[1], Hamed Farahmand[1,*], Russell Blessing[2], Samuel Brody[2], Ali Mostafavi[1]*

[1]Zachry Department of Civil and Environmental Engineering, Urban Resilience. AI Lab, Texas A&M University, College Station, TX, U.S.A.

[2]Institue for a Disaster Resilient Texas, Department of Marine Sciences, Texas A&M University at Galveston, Galveston, TX, U.S.A.

*corresponding author: hamedfarahmand@tamu.edu

**Abstract**
Despite the criticality of dialysis facilities, limited knowledge exists regarding the extent and inequality of disrupted access caused by weather events. This study uses mobility data in the context of 2017 Hurricane Harvey in Harris County to examine the impact of flooding on access to dialysis centers. We examined access dimensions using multiple static and dynamic metrics. The static metric is the shortest distance from census block groups (CBGs) to the closest dialysis centers during the normal period. Dynamic metrics are derived from the analysis of anonymized and aggregated human mobility data; the dynamic metrics are: 1) *redundancy* (daily unique number of dialysis centers visited by residents of CBGs), 2) *frequency* (daily number of visits to each dialysis centers), and 3) *proximity* (daily visits weighted by distance to dialysis centers). By analyzing fluctuations in redundancy, frequency, and proximity metrics during the flood event, we derived spatiotemporal patterns of access disruptions and inequity in mobility from CBGs to dialysis centers. The results show that: (1) the extent of dependence of CBGs on dialysis centers varies; (2) flooding significantly reduces access redundancy and frequency of dialysis center visits; access disruptions persist for more than one month after the flood event; (3) CBGs with a greater minority percentage and lower household income were more likely to experience more extensive access disruptions; (4) high-income CBGs more quickly revert to their pre-disaster levels; (5) larger dialysis centers located in non-flooded areas are critical to absorbing the unmet demand from disrupted facilities; these larger facilities contribute to the absorptive capacity of the network of dialysis facilities in the region. The theoretical contributions of these findings are twofold: (1) the findings show that the extent of disrupted access vulnerability is shaped by the baseline patterns of facility dependence, spatial distribution and capacity of facilities, flood exposure of facilities and CBGs, and inundation of road networks; it is thus critical to incorporate human network dynamics based on human mobility data into an examination of vulnerability to access disruptions and inequalities for critical-care facilities; (2) the findings reveal the presence of inequality in disrupted access among low-income and minority populations; and (3) the findings emphasize the significant role larger facilities play in the absorptive capacity in the network of facilities to accommodate shifting demand when access is disrupted. These findings also have important implications for public health officials and emergency managers to systematically reduce the vulnerability of access disruption to dialysis centers and other critical-care facilities during extreme weather events.

## Introduction

*Background*
Dialysis centers are critical to the health of persons suffering from renal failure or end-stage renal disease. Disasters can disrupt access to dialysis centers, resulting in life-threatening circumstances for dialysis-dependent patients [1]. Lempert and Kopp (2013) defined a kidney failure disaster as "*an event that places* *large number of patients treated with maintenance dialysis or individuals with a recent onset of acute kidney injury (AKI) at risk due to lack of access to dialysis care*" [2]. Extreme weather events, such as hurricanes, flooding, and winter storms, can cause kidney failure disasters due to disrupted access to dialysis centers. Multiple studies have attributed kidney failure disasters to extreme weather events, such as Hurricane Katrina in 2005 [2–5] and Hurricane Gustav in 2008 [6]. In addition, missed regular dialysis sessions due to access disruption can cause second-order impacts, such as an increase in visits to other dialysis centers, a heavier burden of receiving more dialysis-dependent patients, and increased emergency department visits [7,8]. Therefore, better understanding of access disruption to dialysis center services is critical for public health officials and disaster managers to design and implement preparedness and resilience strategies to meet the needs of dialysis-dependent patients.

*Disrupted Access to Dialysis Centers during Flooding*
The majority of existing studies related to access to critical care facilities services focus on service access during normal periods. Rosero-Bixby (2004) defined the demand-supply system connecting census population and health facilities using GIS tools, and then developed an accessibility index based on the distance to the nearest health facility to assess spatial access to health care in Costa Rica [9]. Jin et al. (2015) used the census data at residential building cells as origins, and hospitals and clinics as destinations to examine the spatial inequity of access to healthcare facilities in China [10]. The study developed an index of access to healthcare faculties as a function of the travel time from building cells to closest healthcare facility using a least-cost path analysis. Mayaud et al. (2019) also employed the travel time from grid cells disaggregated on census block groups (CBGs) to healthcare facilities to evaluate the access equity to healthcare facilities in Cascadia [11]. The study used OpenTripPlanner, an open-source routing engine called within the travel-time matrix algorithm of Pereira (2020), to optimize and compute the travel time from grid cells of CBGs to the grid cells of healthcare facilities [12]. Existing studies mostly evaluate access to healthcare facilities by evaluating the existing spatial configuration of healthcare centers; not addressed are the dynamic aspects of access, such as the potential changes in the number of visits to facilities due to a disturbance in the system.

The body of research focusing on the impact of disasters on access to dialysis centers is limited. Kaiser et al. (2021) used the flood map of Hurricane Harvey to examine the flood status of dialysis centers in Harris County [13]. Using the flood zone categories defined by Federal Emergency Management Agency (FEMA), the study categorized dialysis centers by their distance to flooded areas. Solely examining flood exposure of dialysis centers, however, does not provide a full picture of the vulnerability of patients whose access to dialysis centers is disrupted; other factors causing disruptions to dialysis center access could be road flooding [14,15]; flooding of the building causing equipment malfunction or requiring building closure [13,16]; and disruptions in communities where dialysis-dependent patients reside [17].

In this study, we address this knowledge gap by analyzing privacy-preserving human mobility data to examine the extent, spatial patterns, and inequality in disrupted access to dialysis centers in the context of 2017 Hurricane Harvey in Houston metro area. Recent advancements in data acquisition and data analytics have provided opportunities to examine the dynamics of human movement during a community-level disruptive events [18,19] and access to essential facilities in disasters [20].

*Study Objective*
In this study, we used aggregated human mobility data to unveil vulnerability and inequality due to disrupted access to dialysis centers. We specified multiple access metrics to examine both static and dynamic dimensions of access, as well as access inequities in the context of 2017 Hurricane Harvey in Harris County (Houston metro area) in Texas. The static dimension of access was examined based on the shortest distance from CBGs to dialysis centers. The dynamic access dimensions were determined daily number of unique dialysis centers visited by CBG residents (redundancy), daily count of visits to dialysis centers (frequency), and daily distances of visits weighted by visits to dialysis centers (proximity). We computed these metrics based on analysis of the human visitation network (CBG–dialysis center network) and investigated the fluctuations in dynamic access metrics and further exanimated their spatiotemporal patterns to reveal: (1) the extent of dependency of different CBGs on dialysis centers; (2) the spatial patterns of disrupted access and hotspots of vulnerability; (3) inequalities in disrupted access; and (4) the absorption of unmet demand of patients by other facilities. In the following sections, we first describe the data and methods then present the results.

**Materials and Methods**

Figure 1 illustrates an overview of the research steps. To establish access during the normal period, we measured the shortest distance from the population centers of census block groups (CBGs) to the dialysis centers. In the next step, during the flood period, we analyzed the human visitation network from CBGs to different dialysis centers to explore dynamics of access to the dialysis centers. The access dynamics were quantified based on the three access dimensions of redundancy, frequency, and proximity. Then we examined the spatiotemporal patterns of access to identify the extent of dependence of CBGs on dialysis centers and the hotspots of disrupted access. In addition, we examined the absorptive capacity (i.e. the ability of the dialysis centers to maintain unmet needs during a flood event) of the network of dialysis centers by examining the facilities' capacities along with the changes in the patterns of visits during flooding events.

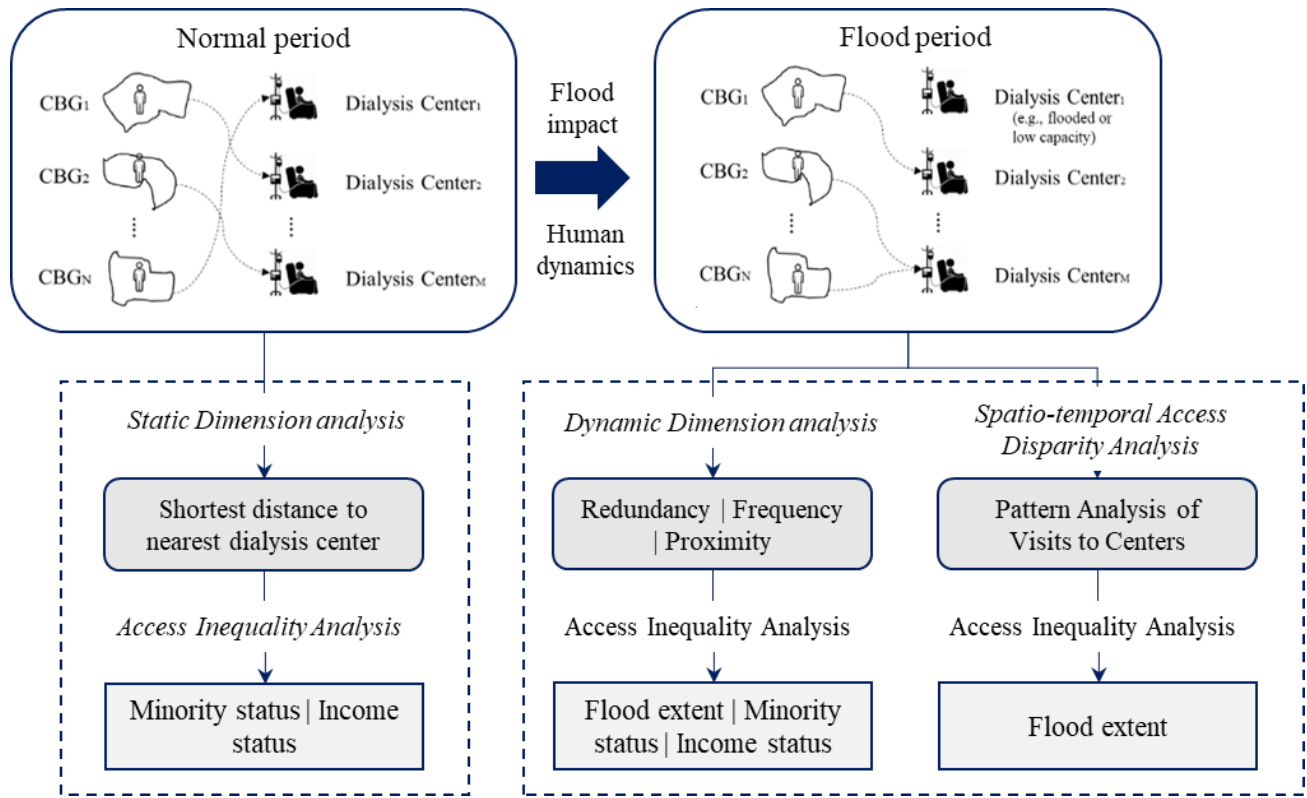

*Figure 1.* Overview of research steps for examining accessibility disruption for dialysis centers due to flood impact. In normal periods, shortest distance to nearest center is established as the static dimension of accessibility. During flooding periods, redundancy, frequency, and proximity metrics are used to examine access disruption and the disparities due to influential factors (i.e., flood extent, minority status, income status) in the census block group scale.

*Study Context*

Our study context is 2017 Hurricane Harvey in Harris County, Texas. Harris County, home to Houston, the fourth largest city in the United States, has seen rapid population growth over past decades. Harris County comprises 2,144 CBGs. Many of CBGs in the county is prone to flood and hurricane hazards due to its coastal location, burgeoning urban development, as well as lack of flood control infrastructure [21]. Harris County also possesses a diverse population with varying sociodemographic characteristics, which provides a representative testbed for exploring the equitable access to dialysis centers across disasters. Within the county are 124 dialysis centers with a total of 2,529 dialysis stations. Hurricane Harvey made landfall in Harris County on August 25, 2017, as a Category 4 storm caused catastrophic impact and wreaked extensive economic and social hardship in Harris County as well as in much of southeast Texas [22]. Due to the hurricane and ensuing flood inundations, nearly 150,000 residents left their flooded homes, and tens of thousands were stranded due to disruptions of roadways [23,24].

*Data Description and Processing*

Multiple datasets were used in this study: location-based human mobility, dialysis center facility data, flood impact data, and demographic data. We specified the study period from August 1, 2017, through September 30, 2017; the study period includes durations of time before, during, and after Hurricane Harvey landfall and the subsequent flooding. Table 1 summarizes the descriptions of each dataset, including data source and provider, processing required for data preparation, variables that have been derived from the processing, and data description.

*Table 1. Data description, processing, and derived variables.*

| Dataset | Description | Processing | Derived variables | Data source |
|---|---|---|---|---|
| Location-based data | This aggregated dataset includes Stop Points by Device. Each row represents a trip from a CBG to different destinations. | 1) Overlapping the polygon of dialysis centers with point data which represents current locations of users; 2) Aggregating number of points within the polygons of dialysis centers by days; 3) Matching each user's 'Place_id' with its 'block_group_id' to denote which trip comes from which CBGs. | Daily trips from CBGs to dialysis centers. | Mobility data provider |
| Dialysis center data | Data on 124 hemodialysis clinics in Harris County, such as location and number of stations. | 1) Filtering locations of dialysis centers; 2) Identifying facility capacities measured by the number of stations. | Locations and dialysis capacity | Medicare.gov |
| Flood Impact data | The flood inundation map of Hurricane Harvey produced by Federal Emergency Management Administration (FEMA). | 1) Overlapping the flood inundation maps of Hurricane Harvey from FEMA with the CBGs map of Harris County; 2) Calculating the flooded percentages of CBGs; 3) Selecting the median of CBGs with flooded area as the threshold to define flood status of CBGs (Flooded: > median of flooded percentage; non-flooded: otherwise); 4) Determining flood status of dialysis centers by projecting their locations within CBGs (Flooded: within flooded CBGs; Non-flooded: otherwise). | Flood statuses of CBGs and dialysis centers | [25] |
| Socio-demographic data | The 2017 5-year estimates data, representing the estimates over the five-year period from 2013 through 2017. | 1) Collecting three variables including the median household income, per capita income, and the ratio of minority (non-white) from the estimates data; 2) Calculating the median of three variables for the 2,144 CBGs in Harris County; 3) Using their medians to denote the levels of minority, household income, and per capita income (Low: <median and High: otherwise). | Levels of minority, household income and per capita income | US Census Bureau |

The anonymized and privacy-enhanced mobile phone data was collected using a Software Development Kit, in which the data provider received informed consent from devices which opted-in to the location data collection. For each anonymous user, more than a hundred data points on average are collected each day, providing an opportunity to gain accurate and precise knowledge of human mobility. High standards of privacy are followed to enable ethical and responsible data collection and use. All data is collected transparently after consent and de-identified. Users are free to opt-out of location sharing at any time. By analyzing the aggregated mobility patterns of more than 500,000 anonymous users (representing 12.5% of the population of the Puget Sound region under analysis), Wang et al. (2019) determined that smartphone-derived GPS data, as compared to cellular network and in-vehicle GPS data, benefits from a superior combination of large-scale, high-accuracy, precision, and observational frequency [26]. Beyond validating scale and accuracy, the research [26] found that the data source is highly demographically representative. The data has a wide set of attributes, including anonymized device ID, point-of-interest ID, latitude, longitude, and dwell time of visitation.

*Methods*

*Access metrics*

As mentioned earlier, we examined multiple access metrics. Table 2 presents the descriptions and equations related to the access metrics used in this study. As can be seen in the Table 2, a shorter distance is more desirable since it generally indicates easier access to service provided by dialysis centers. Redundancy captures the number of unique centers visited by the CBG and higher redundancy means more unique service providers and a higher chance of having access to a dialysis center during a flood. Higher frequency means that the number of daily visits per center is higher, which shows the importance of that facility in terms of demand in case of disruption. Finally, lower proximity captures the average shortest distance for a daily visit to a center.

*Table 2. Descriptions and computation of access dimensions and variables.*

| Variable | Description | Computation |
|---|---|---|
| Shortest distance | Distance from population centers of CBGs to their nearest dialysis centers. | $d_i = min \left\{ 2r \times sin^{-1} \left[ \sqrt{sin^2 \left( \frac{\varphi d_j - \varphi c_i}{2} \right) + cos\varphi d_j \varphi c_i sin^2 \left( \frac{\phi d_j - \phi c_i}{2} \right)} \right] \right\}$ Eq. (1) <br> $r$: earth radius (6,378.137 km); $\varphi d_j$: longitude of dialysis center $j$; $\phi d_j$: latitude of dialysis center $j$; $\varphi c_i$: longitude of the population center of CBG $i$; $\phi c_i$: latitude of the population center of CBG $i$. |
| Redundancy | The daily number of unique dialysis centers visited per CBG on day $d$. | $re_{id} = set(daily\ visited\ dialysis\ centers\ by\ CBG\ i\ on\ day\ d)$ Eq. (2) <br> *set*: function to measure the number of the unique dialysis centers within list of visited dialysis centers per CBG on day $d$. |
| Frequency | The daily number of visits averaged by redundancy per CBG on day $d$. | $fr_{id} = \frac{volume_{id}}{re_{id}}$ Eq. (3) <br> $volume_{id}$: number of daily visits from CBG $i$ to dialysis centers on day $d$. |
| Proximity | The total daily distances from a CBG to all the dialysis centers averaged by redundancy per CBG on day $d$. | $pr_{id} = \frac{\sum_j distance_{ij} volume_{ijd}}{re_{id}}$ Eq. (4) <br> $distance_{ij}$: distance from CBG $i$ to dialysis center $j$; $volume_{ijd}$: number of daily visits from CBG $i$ to dialysis center $j$ on day $d$. |

*Quantification of spatiotemporal access disparities patterns*

Baseline access metrics for access metrics for each CBG were calculated using equations 2-4. The baseline period is August 1, 2017, through August 21, 2017, which captures the time period before any flood related disruptions from Hurricane Harvey had begun (i.e. the normal period). For the remaining days in our study period (i.e., August 22 through September 30, 2017), we computed daily fluctuations for the baseline values of the access metrics using equations 5–7. The fluctuation values are essentially the percent change in each access metric during the flood impact period compared with the baseline period.

$$f\_re_{id} = \frac{re_{id} - re_{ia}}{re_{ia}}$$ Eq. (5)

$$f\_fr_{id} = \frac{fr_{id}-fr_{ia}}{fr_{ia}} \quad \text{Eq. (6)}$$

$$f\_pr_{id} = \frac{pr_{id}-pr_{ia}}{pr_{ia}} \quad \text{Eq. (7)}$$

where, $f\_re_{id}$ is the daily fluctuation between average redundancy and daily redundancy on day $d$; $re_{ia}$ is average redundancy in CBG $i$; $f\_fr_{id}$ is the daily fluctuation between average frequency and daily frequency on day $d$; $fr_{ia}$ is average frequency in CBG $i$; $f\_pr_{id}$ is the daily fluctuation between average proximity and daily proximity on day $d$; $pr_{ia}$ is average proximity in CBG $i$.

To calculate access disruptions, we used the 7-day moving average of percent changes in the access metrics to specify their temporal variation. Next, we implemented an agglomerative clustering algorithm to classify the CBGs into clusters based on their patterns of access disruptions [27]. To do so, the bottom-up approach for clustering started with considering each input as a cluster, and through an iterative greedy process, similar inputs (i.e., clusters) were merged and scaled up the hierarchy. The similarity was calculated based on the Euclidean distance between clusters stored in a proximity matrix. The matrix was updated at each iteration based on the number of clusters and distance between them. Finally, all clusters were merged into one cluster. To implement the algorithm, the number of clusters should be specified; we determined number of clusters by performing a silhouette analysis (Eq. 8). The silhouette coefficient can determine the similarity of a data point within a cluster compared to other clusters. We assessed the number of clusters from one to ten and selected the optimal number of clusters with minimum silhouette coefficient.

$$S(m) = \frac{average\left(\sqrt{(r_N-m)^2}\right) - average_{x \in M}\left(\sqrt{(x-m)^2}\right)}{max\left\{average_{x \in M}\left(\sqrt{(x-m)^2}\right), average\left(\sqrt{(r_N-m)^2}\right)\right\}} \quad \text{Eq. (8)}$$

In Equation 8, $S(m)$ is the silhouette coefficient of the data point $m$, and $m$ is within the cluster $M$; $average_{x \in M}\left(\sqrt{(x-m)^2}\right)$ is the average distance between $m$ and all the other data points in the cluster $M$; $average\left(\sqrt{(r_N-m)^2}\right)$ is the average distance from $m$ to all clusters other than $M$; $r_N$ refers to the centroids of cluster $N$ and $N = 1, 2, \ldots$ (excluding $M$).

## Results and Discussion

*Access Disparities in Normal Period*

We calculated the shortest distance from centroids of CBGs to dialysis centers (Figure 2a). Each polygon represents a CBG, with the shortest distance differentiated by a range of colors. CBGs with the shortest physical distance to dialysis centers (shaded light yellow) are predominantly distributed in the downtown area of Houston. The results also show that the majority of CBGs have access to a dialysis center within 5 kilometers. Figure 2b also shows the distribution of shortest distances from centroids of CBGs to dialysis centers.

To examine inequality in disrupted access, we first evaluated whether a disparity in access exists based on the distance to facilities in the normal period. To explore the access disparities in normal period, we used the sociodemographic attributes of the CBGs—minority population ratio, per capita income, and median of household income—from the U.S. Census 2017 American Community Survey 5-year estimates data for Harris County. Using the median for these three attributes, we defined two income categories. For each variable, CBGs with values higher than median fall in high group; CBGs with values lower than median fall in low group. Figure 2c (from left to right) shows the distribution of shortest distances by minority, per capita income, and median household income. To evaluate the differences in shortest distance distributions across different sociodemographic groups, we conducted the two-sample t-test for these three sociodemographic variables. This analysis evaluates whether high and low groups have significant differences in terms of shortest paths to dialysis centers. Table 3 shows there is no significant difference between CBGs with low and high minority populations. On the other hand, shortest distance of CBGs with low per capita income and low median household income are significantly different from CBGs with high per capita income and high median household income. Patients residing in higher income CBGs have longer shortest path to the nearest dialysis center. Figure 2c also shows that CBGs with either low per capita income or low median household income have less median shortest distances than those with higher incomes. This result implies that the income level as a function of physical distance to dialysis centers may not be a factor contributing to inequality in disrupted access during the flooding period. Figure 2d also shows the dependence of CBGs on dialysis centers based on weekly visits to different centers. CBGs that have visit to higher centers show more dependence to dialysis centers. A large number of the CBGs with higher dependence on dialysis centers are close to downtown in the central region of the Houston metro area.

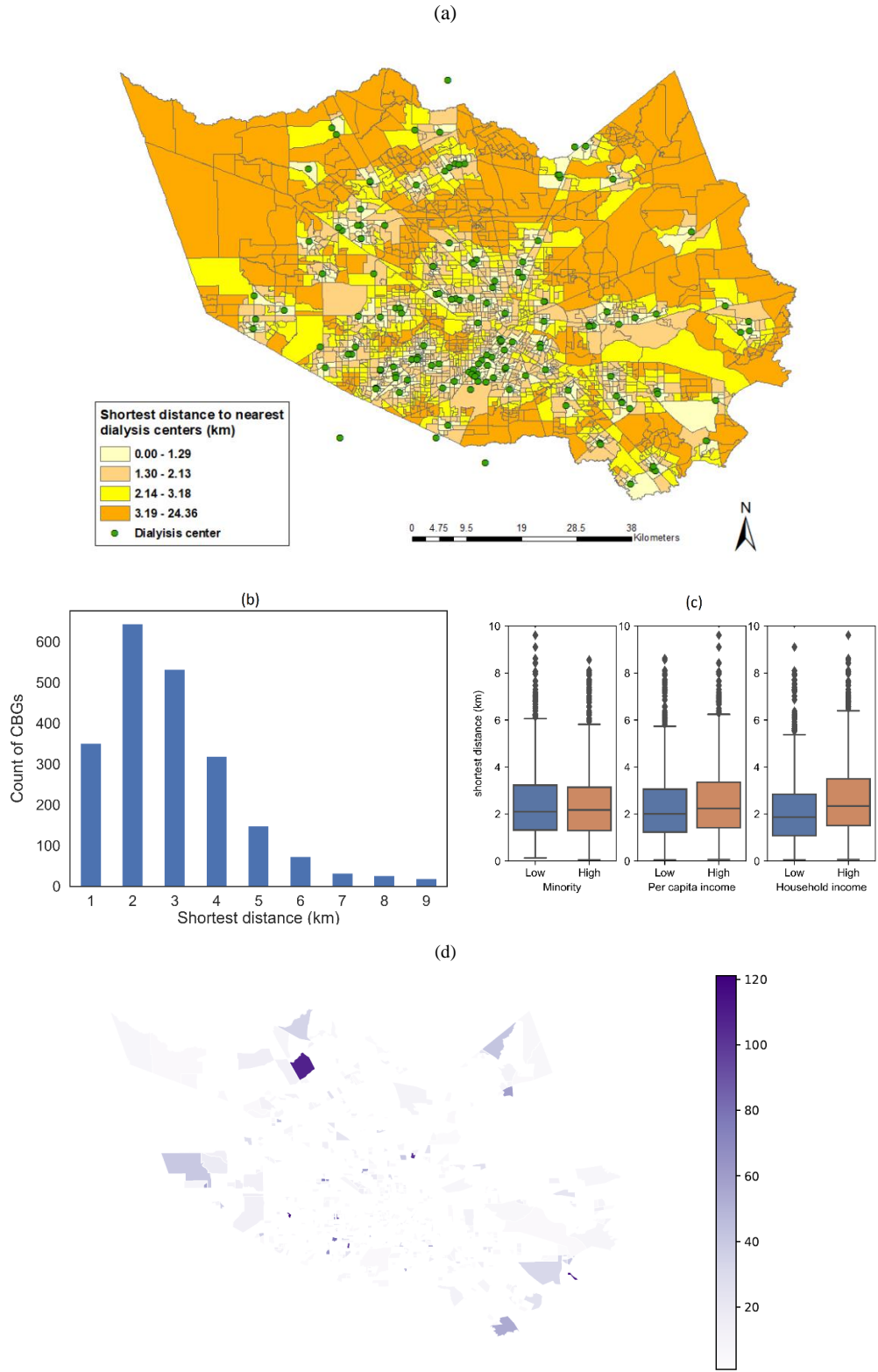

*Figure 2.* Access disparities based on the shortest distance from CBGs to their nearby dialysis centers. (a) spatial distribution of dialysis centers and CBGs with their shortest distances in kilometers (km); (b) count of CBGs in terms of the range of their shortest distances to dialysis centers; (c) sociodemographic characteristics of CBGs versus their shortest distances; (d) dependence on dialysis centers based on weekly visits for different CBGs.

*Table 3. t-test results for shortest distances among different sociodemographic groups.*

| Sociodemographic characteristic | Statistic | p-value |
|---|---|---|
| Minority | -0.80 | 0.42 |
| Per capita income | 3.39 | 0.00069 |
| Median household income | 6.89 | 6.98e-12 |

*Spatial Patterns of Access Disruption*

We used the agglomerative clustering algorithm to analyze CBG-level disparities in dialysis center access disruption. The analysis resulted in four clusters with spatiotemporal patterns of variations in their access metric. Figure 3 and Figure 4 illustrates the results of the clustering analysis. Descriptions and interpretations of the spatiotemporal patterns for access disruptions related to each metric are summarized in Table 4.

As shown in Figure 3 and Figure 4, cluster 1 across all metrics experienced significant access disruption and did not recover until the end of September 2017. It should be noted that a significant access disruption is associated with significant drop in redundancy and frequency as well as surge in proximity. Across all the three metrics, around 80 CBGs fall into cluster 1. The CBGs in cluster 1 for all access metrics are spatially spread out throughout Harris County, indicating there was no spatial effect in the cluster. We can see, however, that some CBGs in cluster 1 are located on the periphery of the county, while the agglomeration of dialysis centers is in the central part of the county. This result implies that centralization of critical care facilities in regions with decentralized population distribution could be a factor in significant access disruptions.

The majority of CBGs fall into Cluster 2 for the three metrics. CBGs in cluster 2 did not experience significant fluctuations in visits, and hence, access disruptions to dialysis centers. Unlike CBGs in cluster 1, CBGs in cluster 2 are located closer to the central part of the county. This result suggests that CBGs located closer to the concentration of dialysis facilities around the center of the county were less vulnerable to access disruptions due to being closer to the concentration of facilities. This result suggests that the proximity of a CBG to the closest facility might not be a reliable metric for evaluating access and vulnerability to access disruptions. Instead, being in a proximity of a concentration of facilities (which includes larger facilities as discussed in the next section) contributes to better access and less vulnerability to access disruption.

Cluster 4 across all metrics appears to include CBGs in which the visits to dialysis centers surged after flooding. These CBGs are in proximity to some of the severely flooded CBGs; hence, a possible explanation for the surge in visits is the relocation of patients from flooded CBGs to the CBGs in the proximity. This explanation is supported by patterns of Cluster 4 in the proximity metric. The distance to dialysis centers for CBGs in cluster 4 of the proximity metric increased; in other words, patients' relocation from flooded areas put them in a location more distant from the dialysis centers they normally visit. While this interpretation is viable, we could not fully validate this interpretation since our mobility dataset was aggregated at the CBG level and we do not analyze individual users' movements. CBGS in cluster 3 show reduction in the proximity metric. By closely looking at the CBGs in this cluster, we find that most of these CBGs are in cluster 2 of the frequency metrics, which indicates that frequency of visits in these CBGs did not change significantly. By juxtaposing the results, we can find that, for the proximity metric, patients in CBGs of cluster 3 chose to go the closest dialysis center to their area to avoid flood-related disruptions (e.g., traffic congestions and busier facilities).

Another noteworthy finding is the persistence of access disruptions in CBGs with the greatest impact until one month after flooding. This result is important since it reveals that disrupted access is not a transitory impact which diminishes along with the recession of flood water. Overall, these results paint a more complete picture of the extent of vulnerability to disrupted access and identifies clusters of CBGs with the greatest access disruption. Further examination of these results reveals the presence of inequality in access disruptions, which is discussed next.

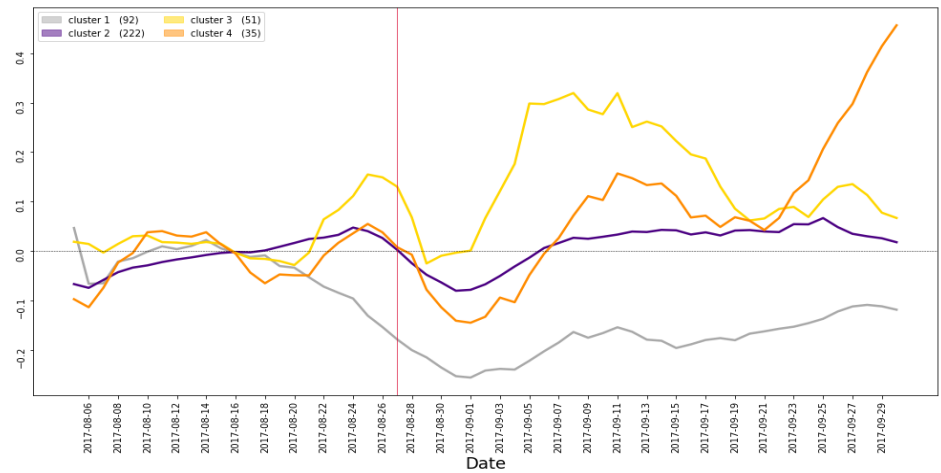

(a)

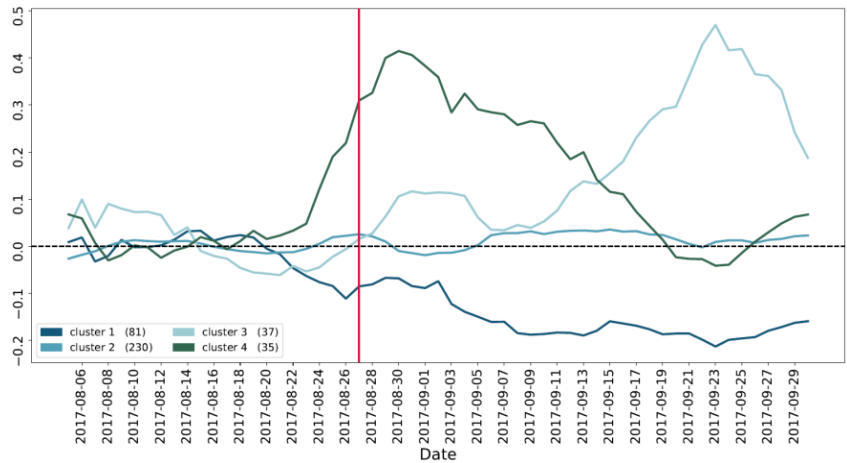

(b)

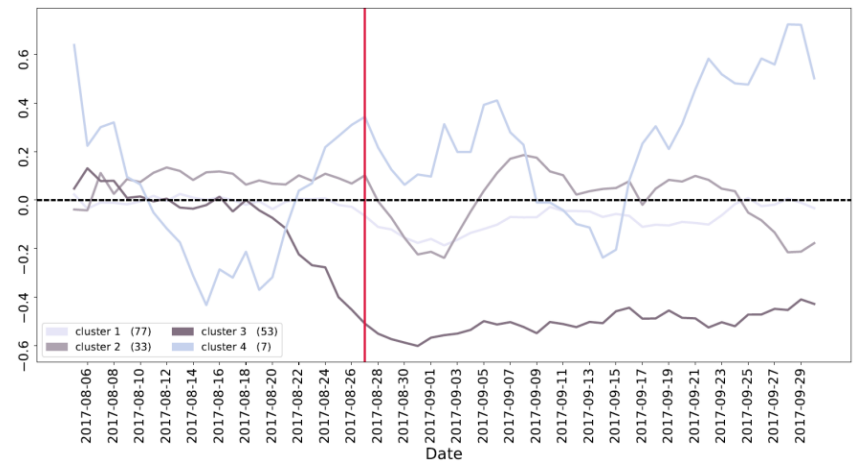

(c)

*Figure 3. Spatiotemporal patterns of dynamic access to dialysis centers during Hurricane Harvey. (a–c) temporal pattern of fluctuations of redundancy, frequency, and proximity; red lines mark the time when Hurricane Harvey made landfall in Harris County; horizontal dashed lines represent the baseline. Y-axis shows the access metric in percentage. It is worth mentioning that CBGs in the same clusters of different access dimensions are not necessarily the same.*

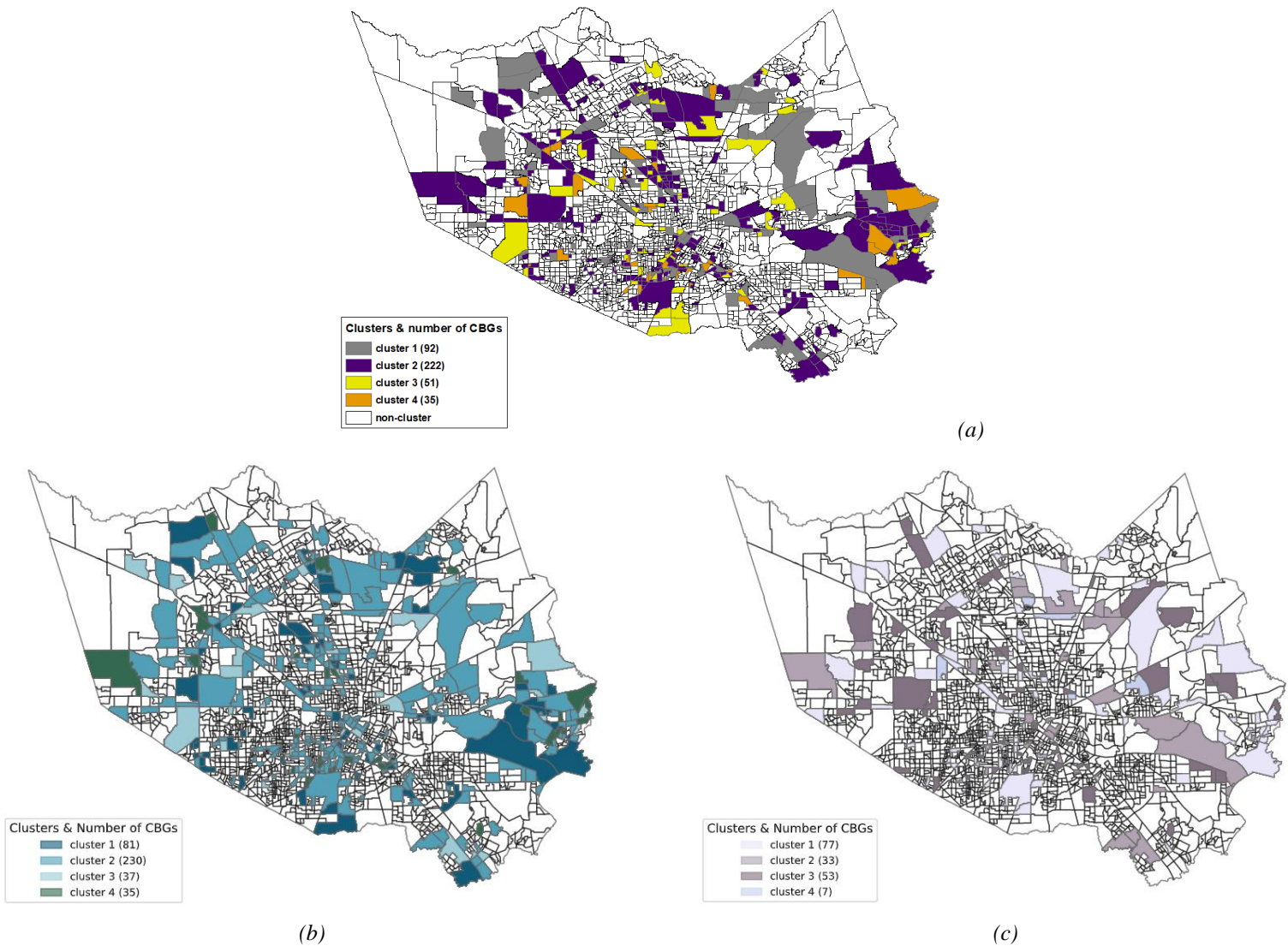

*Figure 4.* Spatiotemporal patterns of dynamic access to dialysis centers during Hurricane Harvey. (a–c) geographic distribution of temporal patterns of variations of redundancy, frequency, and proximity

*Table 4. Descriptions and interpretations of spatiotemporal patterns of dynamic access dimensions.*

| Variable | Cluster | Description | Interpretation |
| --- | --- | --- | --- |
| Redundancy | 1 | Significant decrease of redundancy after Hurricane Harvey and slight recovery by the end September. | Significantly less redundancy compared with baseline means severe disaster impact to access to dialysis centers (fewer options of dialysis centers to visit). |
| | 2 | Decrease of redundancy after Hurricane Harvey and recovery to baseline at the end of first week in September. | Redundancy near the baseline means normal recovery of access activity to dialysis centers (similar number of options of dialysis centers to visit). |
| | 3 | More significant decrease of redundancy after Hurricane Harvey and greater recovery to baseline at the end of first week in September, compared with cluster 2. | Increased redundancy compared to baseline means greater recovery of access activity to dialysis centers (more options of dialysis centers to visit). |
| | 4 | Slight decrease in redundancy after Hurricane Harvey and quick recovery to baseline at the beginning of September. | Non-significantly lower redundancy compared with baseline means slight disaster impact, while quick recovery of redundancy and increased redundancy after hurricane indicate quick and greater recovery of access to dialysis centers. |
| Frequency | 1 | Decrease of frequency after Hurricane Harvey and slight recovery by the end September. | Significantly lower frequency compared with baseline means severe disaster impact to access to dialysis centers (less visits to dialysis centers). |
| | 2 | Slight fluctuations of frequency around the baseline. | Slight fluctuations of frequency mean slight disaster impact on visits to dialysis centers. |
| | 3 | Increased visits to dialysis centers compared with baseline. | Greater recovery of access activities to dialysis centers (more visits to dialysis centers). |
| | 4 | Increased visits to dialysis centers compared with baseline when hurricane made landfall, while slight decrease trend till the end of September. | Increased frequency at the beginning of hurricane means more medical need probably due to disaster, while decreased trend of frequency later means medical needs become mild and recovery of visits to dialysis centers to the normal baseline level. |
| Proximity | 1 | Decrease of proximity after Hurricane Harvey and recovery to the baseline level in almost two weeks. | Good access to dialysis centers or fewer visits to dialysis centers in the first two weeks of the hurricane. |
| | 2 | Decrease of proximity after Hurricane Harvey and recovery to the baseline level in almost one week. | Good access to dialysis centers or fewer visits to dialysis centers in the first week of the hurricane. |
| | 3 | Decrease of proximity after Hurricane Harvey and no recovery to the baseline by the end of September. | Fewer visits to dialysis centers after Hurricane Harvey. |
| | 4 | Significant fluctuations of proximity to dialysis centers. | This cluster contains only seven CBGs, which could be the reason for high fluctuation. |

*Disparities in Disrupted Access*

We examined the spatiotemporal patterns of disrupted access with consideration of sociodemographic characteristics. We used the percent of Hispanic population as the minority status indicator, average household income as the indicator of income status, and total population as an indicator of the population size. Figure 5 shows the boxplots of distribution of sociodemographic characteristics for clusters of CBGs based on patterns of disruption of access based on redundancy, frequency, and proximity. Using the insights from the results in Figure 5 together with spatial and temporal patterns of access disruptions found in Figure 3, we investigated the presence of inequality in disrupted access to dialysis centers. To do so, we compared the distribution of sociodemographic characteristics of clusters with highest access disruption (cluster 1 and cluster 2) with other clusters. As can be seen in Figures 5a and 5b, cluster 1 and cluster 2 have higher ratios of minority populations. This result shows that areas with minority status experienced more severe access disruptions to dialysis centers during the flood event, especially in terms of redundancy and frequency indicators.

Figure 5d–f shows the association between population size and access disruption extent. As can be seen, there is no considerable difference in population size among different clusters. Figures 4g–i shows the association between income level and experienced disruption of access to dialysis centers indicated by cluster numbers. Exploring the trend of access redundancy shows that clusters 4 and 3, which experienced considerably lower access disruption compared to clusters 1 and 2, have higher household income. Similarly, a general increasing trend can be observed in Figure 4i that shows that as the disruption extent decreases from cluster 1 to cluster 4, household income increases. However, no distinctive trend can be observed in the visit proximity–income relationship. Therefore, it can be concluded that the areas with higher income had higher redundancy and frequency of access to dialysis center that provide better access to dialysis services during and in the immediate aftermath of the flooding compared with areas with lower income.

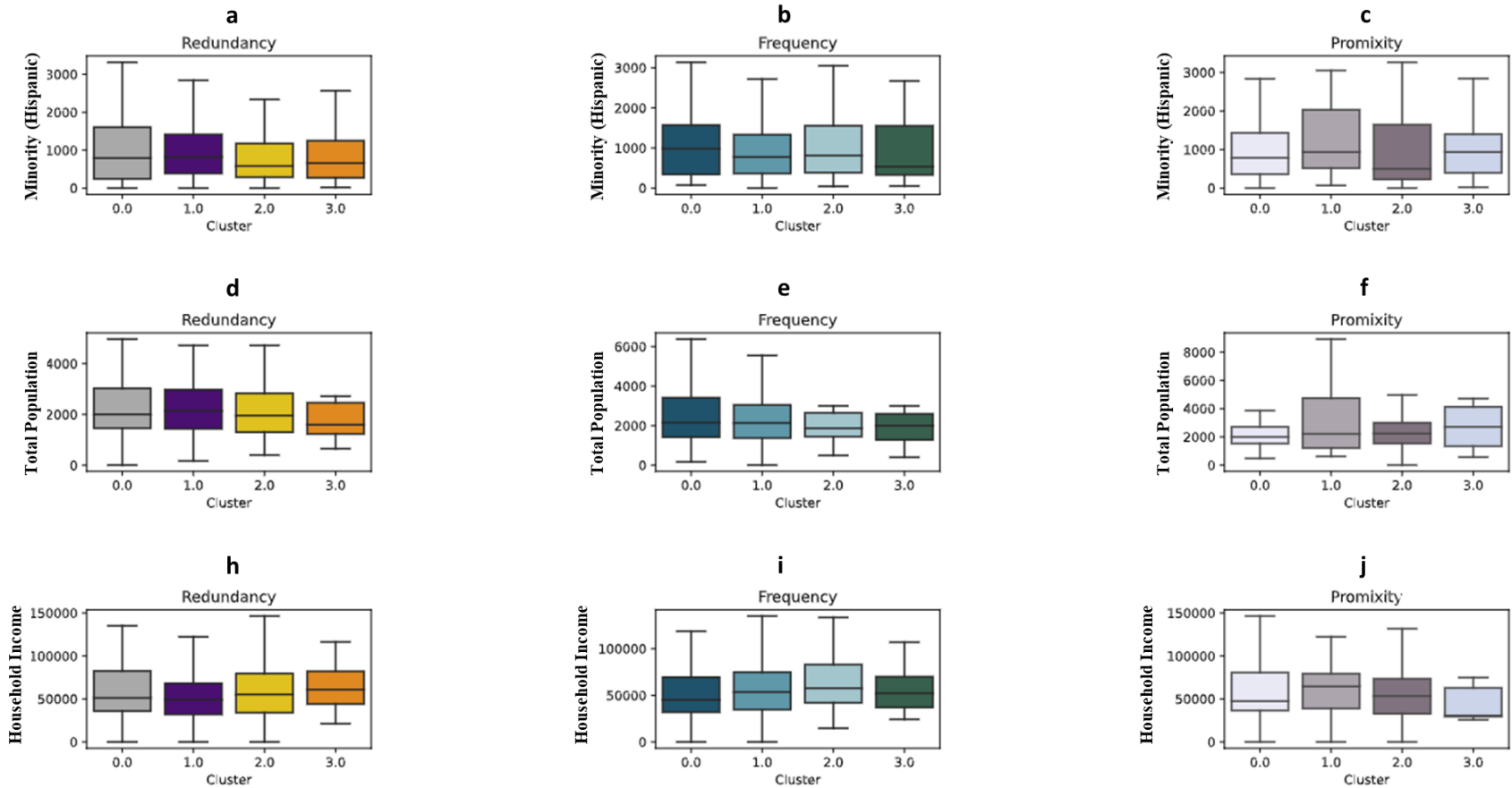

*Figure 5.* *Sociodemographic characteristics of identified spatiotemporal clusters related to fluctuations in access to dialysis centers. (a-c) minority status distribution of different clusters for each access indicators; (d-f) total population distribution of different clusters for each access indicators (g-i) household income distribution of different clusters for each access indicators.*

*Varying Services Capacity Provided by Dialysis Centers*

In the next step, we quantified the service levels provided by dialysis centers based upon the number of CBGs that visit facility. Using an agglomerative clustering algorithm, setting boundaries with a silhouette coefficient, we identified three clusters of dialysis centers based on service levels (Figure 6). Most of the dialysis centers were in cluster 1. Facilities in this cluster received patients from only two to three CBGs. The number of CBGs that visited the facilities in cluster 1 did not show significant fluctuation during the flooding period. Dialysis centers in cluster 2 served a greater number of CBGs during normal period than clusters 1 or 3. While the number of CBGs which visited the facilities in cluster 2 decreased in the aftermath of Harvey, the numbers returned to the normal period within a week. Dialysis centers in cluster 3 showed a medium level of service based on the number of CBGs they serve. They experienced a slight decrease in the number of CBGs that visited and recovered to their previous service level in almost two weeks.

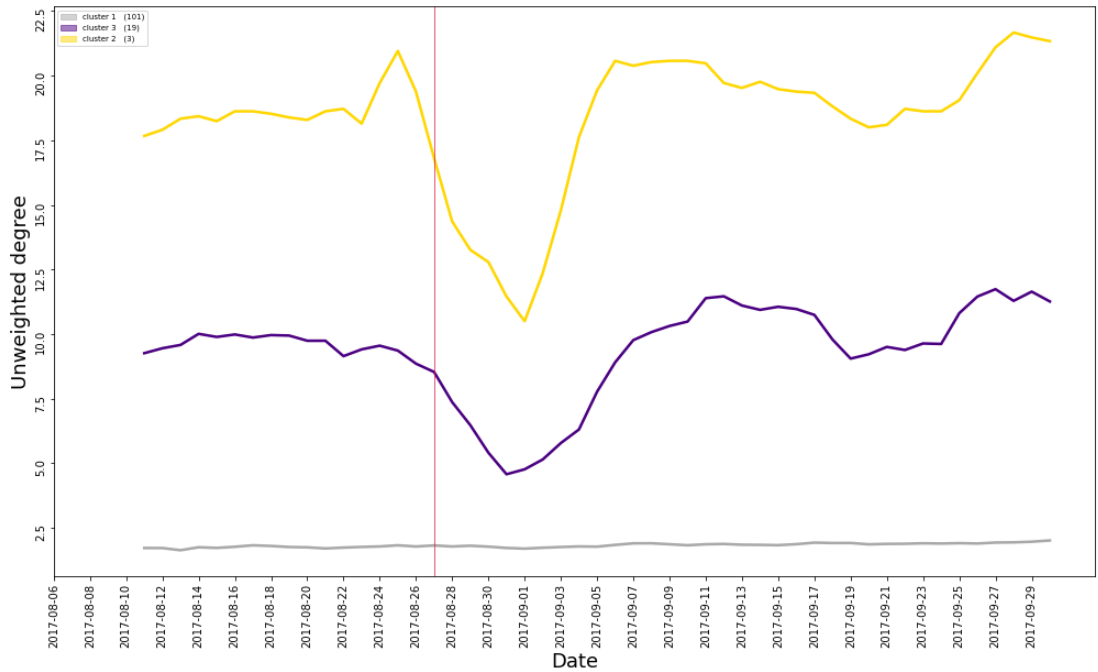

(a)

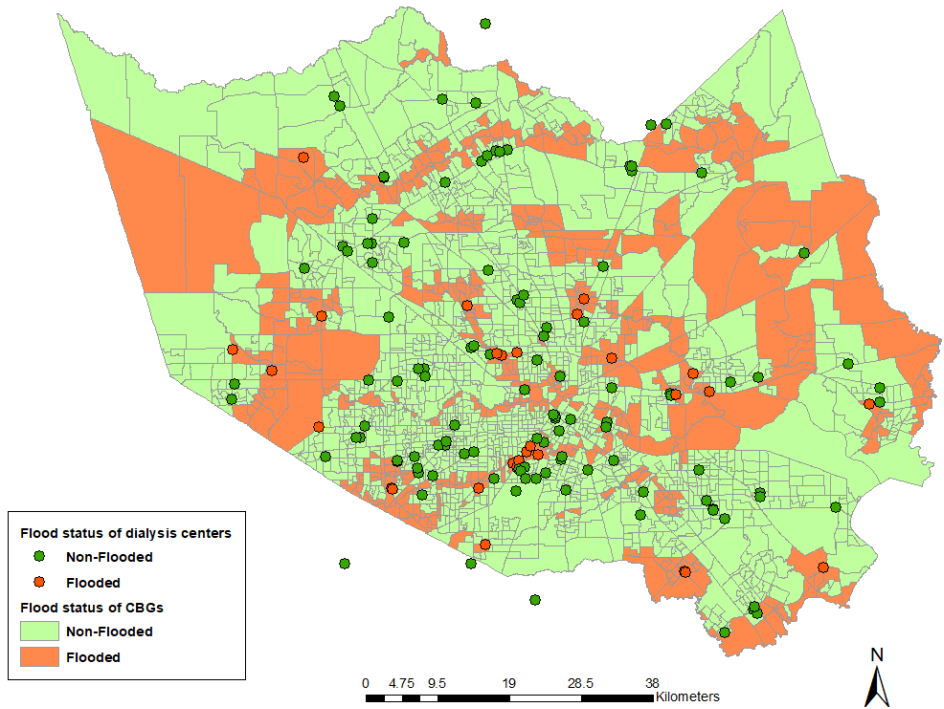

(b)

*Figure 6.* *Spatiotemporal pattern of provided services levels of dialysis centers (a) and geographic distribution of the identified clusters and flooded status of CBGs in Harris County (b).*

To further explore variations in the provided service levels by dialysis centers, we considered three factors:

(1) *flood inundation status of dialysis centers:* if the dialysis center is in an inundated CBG, it is denoted as flooded (although the facility might not have flooded: the rationale is if a facility is in a flooded CBG, accessing the facility would be more difficult due to road inundations, closures, and other disruptions);

(2) *capacity:* we used the number of stations within dialysis centers to represent their service capacities. Using the median of numbers of stations of all the dialysis centers, we defined the threshold as 17 to determine the capacity levels of dialysis centers: centers with more than 17 stations (66 centers) are considered as *high capacity* while other centers are considered as *low capacity* (58 dialysis centers); and

(3) *Combined evaluation of flood inundation status with dialysis center capacity:* we also used the combination of flood inundation status of the center and its capacity. In this analysis, a facility can have four statuses based on the status of flood inundation (flooded/non-flooded) and the capacity of the center (high/low).

The analysis results are shown in Figure 7 According to Figure 6a, we can see the percentage of dialysis centers considered as flooded in cluster 3 is much larger than that in cluster 1. This could explain the fact that dialysis centers in cluster 3 experienced moderate fluctuations in terms of the number of CBGs that visited them, while number of visits from CBGs to the dialysis centers in cluster 1 remained almost constant (i.e., not considerable fluctuations in the number of visits). Furthermore, the capacity of the dialysis centers plays a major role in the variation of visited CBGs. Figure 6b shows a larger percentage of dialysis centers with high capacity in cluster 1 than that in cluster 3. On the other hand, we know that the number of CBGs served by dialysis centers in in cluster 1 is much lower than cluster 3. This result implies that patients' decision regarding visiting a dialysis center is not determined by the capacity of the center, which might be an important factor during emergency situations when the supply decreases due to malfunction of a portion of centers and cause delay for service provision in centers with low capacity and increased demand. This finding shows a need for proper planning for enhancing patient awareness and re-distribution of patients with unmet treatment to other centers based on a system-optimization approach that accounts for both capacity of all facilities, as well as travel distance. This can include providing information about centers with high capacities located in areas with lower risk of flooding due to direct impacts (i.e., facility inundation) or indirect impact (i.e., loss of access due to road closure and loss of functionality due to power outage or supply shortage). Such a patient-center allocation optimization system during emergency events (such as flooding) could increase the overall resilience of the network of facilities. To further explore the fluctuation in visits between low- and high-capacity centers, we plotted the distribution of maximum percentage change of visits in flood period compared with normal period for centers of high and low capacity. As the boxplot Figure 6d shows, the general shape and median of the distributions are similar, while the low-capacity group has slightly higher maximum percentage change inclined more toward positive changes (i.e., increase in visits). In fact, higher percentage increase has been observed in low-capacity centers, which highlights the importance of both low- and high-capacity dialysis centers in absorbing the unmet demand of the patients.

With the combined analysis of the impacts of flood status and service capacity, we found that the proportion of dialysis centers with high capacity and non-flooded status is much larger than the sum proportions of those with high/low capacity and flooded status; this result can explain the absence of significant fluctuation in the provided service level of dialysis centers in cluster 1. In cluster 3, the number of dialysis centers with high capacity and flooded status is larger than those with high capacity and non-flooded status, which resulted in the moderate fluctuation in the provided service levels of dialysis centers in this cluster. These results suggest that larger capacity centers that, assuming if they remain accessible, can absorb the unmet demand of patients from other dialysis centers and contribute to the absorptive capacity of the network of facilities.

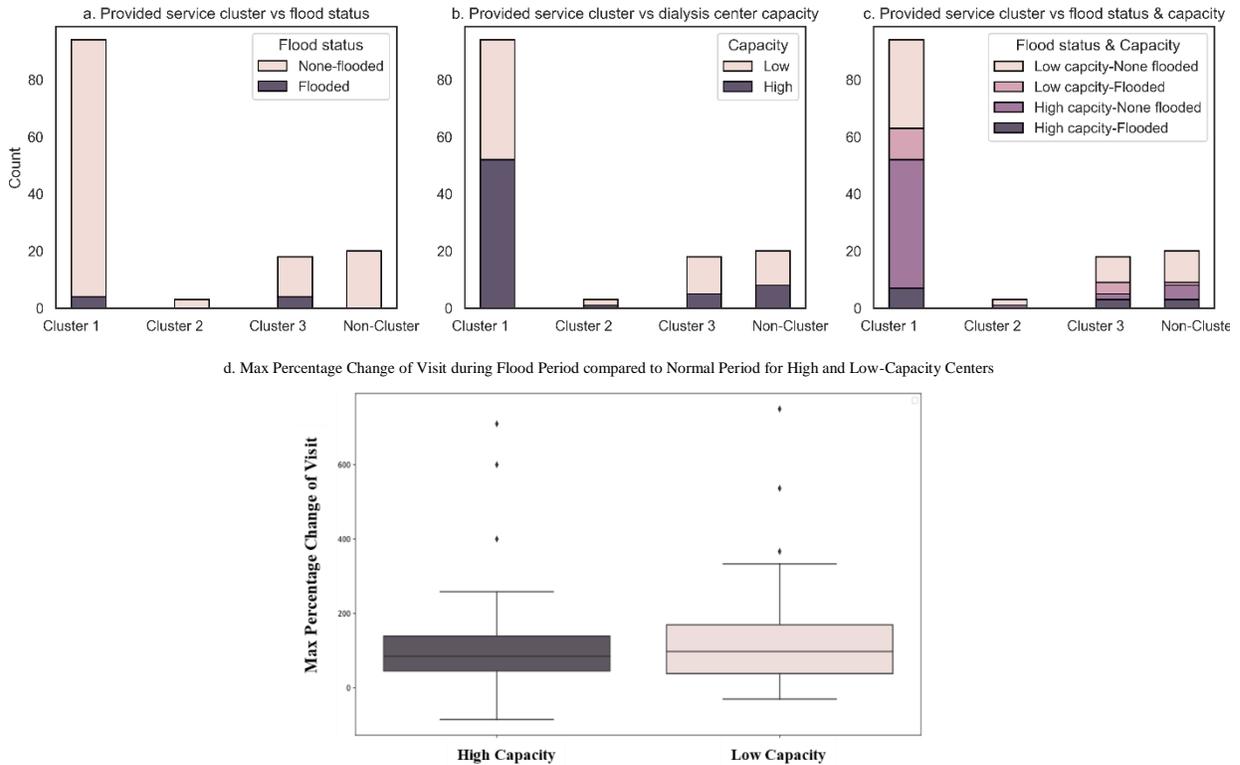

*Figure 7.* Flood inundation status (a), capacity (b), and integration of flood status and capacity (c) of different clusters of dialysis centers based the number of daily visited CBGs; (d) maximum percentage change of visits to high- and low-capacity dialysis centers during flood period compared to normal period.

## Conclusions

### Summary of findings

Disruption in dialysis service provision could have life-threatening implications for dialysis-dependent persons constituting a medically vulnerable population [28,29]. Natural hazards such as floods can impact access to dialysis centers through disruption of access to the center (i.e., access road closure), disruptions that limit patients' ability to access to the service (i.e., flood inundation in the neighborhood), and closure of dialysis centers. Therefore, it is crucial to examine access disruption to dialysis centers during both normal conditions and during flooding periods to meet the medical needs of persons this medically vulnerable population. To address the lack of analysis of access disruption to dialysis centers during floods, we evaluated the accessibility disruption due to flood impact on dialysis centers in the context of 2017 Hurricane Harvey in the Houston metro area and broader Harris County, Texas. We defined metrics to capture static and dynamic access dimensions. Shortest distance from CBGs to closest dialysis centers during normal period were used as the main indicator of accessibility of different CBGs during the normal period. To capture fluctuations in accessibility during the flood period and to quantify the access dynamic, we analyzed anonymized, aggregated human mobility data. Using this data, we examined three dynamic metrics including: 1) redundancy (i.e., daily unique number of dialysis centers visited by residents of CBGs), 2) frequency (i.e., daily number of visits to each dialysis center), and 3) proximity (i.e., daily visits weighted by distance to dialysis centers). We analyzed fluctuations in redundancy, frequency, and proximity metrics during the flood event and explored the spatial patterns of areas with similar access disruption based on the fluctuation of different metrics.

The analysis revealed important findings regarding the impact of flood on the access to dialysis centers and the underlying mechanisms and disparities in the fluctuations in visits to dialysis centers. (1) the extent to which each CBG is dependent upon dialysis centers varies and it can be quantified with the developed metrics; (2) the redundancy and frequency of access to dialysis centers is significantly reduced due to flood disruption and can persist for more than one month after the flood event; (3) areas with a larger ratio of minority population and lower household income experienced a greater disruption in access to dialysis centers; (4) access patterns of high-income CBGs recover more quickly to their pre-disaster levels; (5) dialysis centers with high capacity that are located in areas with lower exposure to flood inundation are critical to absorb the unmet demand from disrupted facilities and thus can be considered as facilities that contribute to the absorptive capacity of the network of dialysis facilities in the region.

Methodologically, this research demonstrates how human dynamics of large-scale mobility data can be used as a tool to examine the access disparities to dialysis centers across normal and flood periods. To quantify the access to dialysis centers, we defined a four-dimension access metric for both normal (static) and disaster (dynamic) periods. For fluctuations of dynamic access dimensions, we implemented an agglomerative clustering algorithm to identify the spatiotemporal patterns of their access disparities to dialysis centers. Our proposed framework for examining access inequity issues from CBGs to dialysis centers can be directly applied to the access studies of other critical facilities, such as groceries and drugstores. In addition, this study contributes to the body of knowledge of smart flood resilience field [30]. One essential component of smart flood resilience is predictive infrastructure failure monitoring, which considers the assessment of neighborhood vulnerability to lose access to hospitals. With our defined dynamic access dimensions to dialysis centers derived from by human dynamics, future research can include access disparities to essential facilities into the smart flood resilience framework.

The theoretical contributions of this study are twofold: (1) findings revealed that baseline patterns of facility dependence, spatial distribution and capacity of facilities, flood exposure of facilities and CBGs, and inundation of road networks are major determinants of the extent of access disruption. Hence, it is critical to incorporate human network dynamics based on human mobility data in examining vulnerability of access disruptions and inequalities for critical care facilities; and (2) the analysis showed that inequalities in access redundancy, frequency, and proximity among low-income and minority populations need to be addressed. In addition, in terms of the emergency management and public health planning for dialysis centers, the findings indicate that the role of the absorptive capacity of high-capacity centers located in areas with lower inundation exposure should be incorporated into planning for dialysis-dependent patients. This can significantly facilitate accommodating unmet demands of facilities to which access is disrupted. The approach to analyzing access disruption to dialysis centers and the findings have important implications for public health officials and emergency managers to systematically examine the vulnerability of access disruption to dialysis centers and other critical care facilities during extreme weather events to inform their preparedness and resilience plans.

**Acknowledgments**

The authors would like to acknowledge funding support from the Texas A&M X-Grant Presidential Excellence Fund, as well as the National Science Foundation CRISP 2.0 Type 2 No. 1832662. Any opinions, findings, conclusions, or recommendations expressed in this material are those of the authors and do not necessarily reflect the views of the National Science Foundation.

**Data availability**

All data were collected through a CCPA- and GDPR-compliant framework and utilized for research purposes. The data that support the findings of this study are available from the data provider, but restrictions apply to the availability of these data, which were used under license for the current study. The data can be accessed upon request from the data provider. Other data we use in this study are all publicly available.